\documentclass[aps,prx,reprint,preprintnumbers,superscriptaddress,nofootinbib,longbibliography,floatfix,onecolumn]{revtex4-2}
\pdfoutput=1
\usepackage{rotating}
\usepackage{array}
\usepackage[normalem]{ulem}
\usepackage{slashed}
\usepackage{booktabs}
\usepackage[pdftex,table]{xcolor}
\usepackage{units}
\usepackage{xfrac}
\usepackage{mathtools}
\usepackage{empheq}
\usepackage{multirow, booktabs}
\usepackage{amssymb, amsmath}
\usepackage{url}
\usepackage{comment}
\usepackage{physics}
\usepackage{color,soul}
\usepackage{bbm}
\usepackage[caption=false]{subfig}
\usepackage[linesnumbered,ruled,vlined]{algorithm2e}

\definecolor{darkred}{rgb}{1.0,0.1,0.1}
\definecolor{darkgreen}{rgb}{.01, .66, .32}

\usepackage{hyperref}
\hypersetup{
  colorlinks=true,
  citecolor=blue,
  linkcolor=blue,
  urlcolor=blue
}

\usepackage{lineno}

\begin{document}

\title{Neural Posterior Unfolding}

\author{Fernando Torales Acosta}
\affiliation{Physics Division, Lawrence Berkeley National Laboratory, Berkeley, CA 94720, USA}
\affiliation{Department of Physics, University of California, Berkeley, CA 94720, USA}

\author{Jay Chan}
\affiliation{Scientific Data Division, Lawrence Berkeley National Laboratory, Berkeley, CA 94720, USA}

\author{Krish Desai}
\affiliation{Physics Division, Lawrence Berkeley National Laboratory, Berkeley, CA 94720, USA}
\affiliation{Department of Physics, University of California, Berkeley, CA 94720, USA}

\author{Vinicius Mikuni}
\affiliation{National Energy Research Scientific Computing Center, Berkeley Lab, Berkeley, CA 94720, USA}

\author{Benjamin Nachman}
\affiliation{Fundamental Physics Directorate, SLAC National Accelerator Laboratory,
Menlo Park, CA 94025, USA}
\affiliation{Department of Particle Physics and Astrophysics, Stanford University,
Stanford, CA 94305, USA}

\author{Jingjing Pan}
\email{jingjing.pan@yale.edu}
\affiliation{Department of Physics, Yale University, New Haven, CT 06511, USA}
\affiliation{Physics Division, Lawrence Berkeley National Laboratory, Berkeley, CA 94720, USA}

\author{Francesco Rubbo}
\affiliation{Altos Labs, Redwood City, California, 94065, USA}

\begin{abstract}
    Differential cross section measurements are the currency of scientific exchange in particle and nuclear physics.
    A key challenge for these analyses is the correction for detector distortions, known as deconvolution or unfolding.
     Binned unfolding of cross section measurements traditionally rely on the regularized inversion of the response matrix that represents the detector response, mapping pre-detector (`particle level') observables to post-detector (`detector level') observables.
     In this paper we introduce Neural Posterior Unfolding, a modern, Bayesian approach that leverages normalizing flows for unfolding.
     By using normalizing flows for neural posterior estimation, NPU offers several key advantages including implicit regularization through the neural network architecture, fast amortized inference that eliminates the need for repeated retraining, and direct access to the full uncertainty in the unfolded result.
     In addition to introducing NPU, we implement a classical Bayesian unfolding method called Fully Bayesian Unfolding (FBU) in modern Python so it can also be studied.  These tools are validated on simple Gaussian examples and then tested on simulated jet substructure examples from the Large Hadron Collider (LHC).  We find that the Bayesian methods are effective and worth additional development to be analysis ready for cross section measurements at the LHC and beyond.
\end{abstract}

\maketitle


\section{Introduction}
\label{sec:intro}

Cross section measurements form the critical interface between theoretical predictions and experimental observations in particle and nuclear physics.
Given some observables (what data scientists might call `features'), theorists attempt to predict and experimentalists measure how often a given reaction produces a certain value of the observable. Together, they use these inputs to infer properties of particle and nuclear physics models.
In practice, theoretical calculations produce predictions at the particle level---the intrinsic, pre--detector distributions of physical observables---while experiments record data at the detector level, where measurements are distorted by detector effects such as resolution and acceptance limits and inefficiencies.
Unfolding (also known as deconvolution) is the process of reconstructing the true particle level distributions from these distorted detector-level observations.

Thousands of unfolded cross section measurements have been performed, nearly all of them using classical methods~\cite{Adye:2011gm}.  Most of these methods discretize the problem, representing the differential cross section as a histogram.
While there is a growing interest in using machine learning for unbinned unfolding~\cite{Arratia:2021otl,Huetsch:2024quz}, we focus here on the (currently) more prevalent binned methods.
Frequentist methods that employ maximum likelihood estimation (and its approximations) for unfolding dominate particle and nuclear physics analyses~\cite{Cowan:2002in,Blobel:2203257,doi:10.1002/9783527653416.ch6,Balasubramanian:2019itp}.
Directly maximizing the likelihood, however, is numerically unstable, especially when detector distortions are non-negligible, because the detector response kernel is highly singular. As a result, all unfolding methods perform a variation of regularized maximum likelihood estimation to prevent small fluctuations in the the detector-level data leading to large changes in the predicted particle-level distributions.

An alternative approach is to use Bayesian methods with a uniform prior.  Fully Bayesian Unfolding~\cite{choudalakis2012fully} has been used for a number of measurements~\cite{ATLAS:2013buu,ATLAS:2015ysm,LHCf:2015rcj,ATLAS:2015jgj,ATLAS:2015sex,ATLAS:2016ykb,ATLAS:2016rnf,ATLAS:2016bac,PHENIX:2019pxh,ATLAS:2021dqb,ATLAS:2022uav,ATLAS:2022waa}, with a uniform prior and Markov Chain Monte Carlo (MCMC).
Many of these measurements are based on pyFBU\footnote{\url{https://github.com/pyFBU/}}, a decade-old, no longer maintained library, and deprecated software including Theano~\cite{thetheanodevelopmentteam2016theanopythonframeworkfast}.
%
%
One of our goals is to provide an updated version of Fully Bayesian Unfolding that makes use of current tools.
We also explore an alternative approach to MCMC by using normalizing flows~\cite{rezende2016variational} and amortized machine learning. 
The proposed method uses pairs of (truth, measured) histograms to learn $p(t | m)$, the probability density function of truth given measured.\footnote{Note that the causal structure is in the opposite direction.}.
Using this approach rather than MCMC offers the potential advantages of providing a function approximation of the likelihood and eliminating the need to rerun the algorithm when certain changes are applied to the data (e.g. addition of more data or determination statistical uncertainties with bootstrapping~\cite{10.1214/aos/1176344552}).  Furthermore, as a neural network, normalizing flows are inherently regularized through the limited complexity of the architecture and training protocol.  We call this new method  \emph{Neural Posterior Unfolding} (NPU).

%
%
%
%
%
%
%

The remainder of the paper is organized as follows.
Section~\ref{sec:stats} reviews the statistics of binned unfolding and outlines the challenges the inverse problem presents.
Sec.~\ref{sec:ml} introduces and details the implementation of NPU using modern tools.
Section~\ref{sec:results} presents numerical results from both Gaussian and simulated high-energy physics datasets, and Section~\ref{sec:conclusions} concludes with a discussion of the advantages, limitations, and future research directions of our approach.

\section{Statistics of Unfolding}
\label{sec:stats}

For binned unfolding, the task is to infer the cross section $t_i$ in each `particle level' (pre--detector) bin $i$ given the observed counts $m_j$ in each `detector-level' (post-detector) bin $j$ and response matrix with elements $R_{ij}=\Pr(m_i|t_j)$, estimated from simulation.
When the response matrix is correctly modeled (a necessary assumption), on average, $\mathbf{m = R\, t}$.
The naive approach is to analytically invert the response matrix, $\mathbf{\hat{t}=R^{-1}\,m}$.
While unbiased, this approach suffers from two major limitations.
First, the response matrix modelled from simulations might not be non-negative definite, which can yield negative predictions for \(t_i\).
This would violate the physical requirement that cross sections be non--negative.
Second, matrix inversion amplifies statistical fluctuations due to (often large) off-diagonal components of $\mathbf R$.
This can result in small perturbations in $\mathbf{m}$ causing large perturbations in $\mathbf{\hat{t}}$.  The matrix itself may also not be square so analytical inversion might not be possible altogether.

To address these challenges, a number of regularized matrix inversion solutions have been proposed.
One of the most widely used approach is Iterative Bayesian Unfolding (IBU) (also known as Lucy-Richardson deconvolution)~\cite{DAgostini:1994fjx,Richardson:72,1974AJ.....79..745L}.
Despite its name, IBU is a strictly frequentist approach;
it is an expectation-maximization (EM) technique that converges to a (local) maximum likelihood estimate~\cite{shepp1982maximum}.  
%
%
IBU will serve as our baseline.
The iterative update in IBU is expressed as
\begin{align}
    t_j^{(n)} &= \sum_i\text{Pr}_{n-1}(\text{truth $j$}|\text{measure $i$})\;\Pr(\text{measure $i$})\\\label{eq:ibu}
    &=\sum_i\frac{R_{ij}\,t_j^{(n-1)}}{\sum_k R_{ik}\,t_k^{(n-1)}}\times m_i\,,
\end{align}
for some initial guess $t^{(0)}$ (usually the starting simulation) and up to some maximum $n_{\max}$ iterations.
Cutting of the number of iterations before convergence, $n_{\max}<\infty$, is a form of regularization that prevents overfitting to statistical fluctuations in the data.
IBU only provides a point estimate, but it is possible to estimate statistical uncertainties through asymptotic formulae or through bootstrappig~\cite{Efron_1979}.
This already highlights potential advantages of NPU: the statistical uncertainty is part of the result and the regularization is implicit in the training and automated through model selection (via the validation loss).

Another difference between NPU and IBU is their behavior in unconstrained regions of phase space.
Cross sections in the particle-level phase space that are unconstrained or poorly constrained by detector level measurements should be highly uncertain.
To understand the behavior of the two methods on degenerate data, consider two particle level bins \(\alpha, \beta\) that are indistinguishable at detector level.
Formally, there exists a detector level bin \(\kappa\) such that for every detector level bin \(\iota,\; R_{\iota\alpha} = R_{\iota\beta} = \delta_{\iota\kappa}.\)
where \(\delta\) is the Kronecker delta symbol.
In this case, the IBU update rule in Eq.~\ref{eq:ibu} yields:

\begin{align}
&\forall n\in\mathbb{N}\;t_{\alpha}^{(n)}=\frac{t_\alpha^{(n-1)}m_\kappa}{t_\alpha^{(n-1)}+t_\beta^{(n-1)}}\quad\textrm{and}\quad t_{\beta}^{(n)}=\frac{t_\beta^{(n-1)}m_\kappa}{t_\alpha^{(n-1)}+t_\beta^{(n-1)}}\,\\
\implies &\forall n\in \mathbb{N}\;t_{\alpha}^{(n)} = \qty(\frac{t_\alpha^{(0)}}{t_\alpha^{(0)} + t_{\beta}^{(0)}})\,m_\kappa\quad\textrm{and}\quad t_{\beta}^{(n)} = \qty(\frac{t_\beta^{(0)}}{t_\alpha^{(0)} + t_{\beta}^{(0)}})\,m_\kappa\,.
\end{align}
which means that $\forall n\in\mathbb{N}\;$ the reweighting factor is exclusively a function of the prior, and is entirely independent of the data.
IBU thus reports that the uncertainty on $\hat{t}_\alpha/(\hat{t}_\alpha+\hat{t}_\beta)$ is zero, even though it should be maximal.
This uncertainty is currently somewhat covered through systematic uncertainties estimated by comparing different priors, usually with a small set of simulations.
These are not sufficient and in fact the reported uncertainty could still be zero if the two priors happen to agree in these bins.
A similar analysis holds for other regularized matrix inversion techniques.
In contrast, methods like Fully Bayesian Unfolding (FBU) and Neural Posterior Unfolding (NPU) return a full posterior distribution over the unfolded parameters.
This approach naturally captures the inherent uncertainty, especially in regions where the data provide weak constraints, by not forcing a single point estimate.
NPU thus not only mitigates the risk of underestimating uncertainty in degenerate scenarios but also provides a principled framework for uncertainty quantification.  
Classical statistical uncertainties with IBU (e.g. with bootstraps) are able to capture some of these uncertainties, but not in the highly degenerate cases like the one from above.

\section{NPU Machine Learning Implementation}
\label{sec:ml}

We implement Neural Posterior Unfolding (NPU) using  \textsc{tensorflow}~\cite{tensorflow2015-whitepaper} and \textsc{Tensorflow Probability}~\cite{DBLP:journals/corr/abs-1711-10604}.
The normalizing flow implementation is based on the MADE~\cite{germain2015made,papamakarios2018masked} implementation consisting of an invertible transformation network composed of three fully connected layers with $50 - 100$ nodes each, employing the \textsc{Swish}~\cite{ramachandran2017searching} activation function.
Conditional inputs, based on post--detector observables, are incorporated through an auxiliary fully connected layer.
Table~\ref{tab:hyperparameters} summarizes the key hyperparameters used in our experiments.

\begin{table}[h!]
\centering
\caption{Summary of hyperparameters for training the NPU model.}
\label{tab:hyperparameters}
\begin{tabular}{|c|c|}
\hline
\textbf{Hyperparameter} & \textbf{Value}\\
\hline
Number of Layers & 3 (main) + 1 (conditional)\\
Nodes per Layer & 50–100\\
Activation Function & \textsc{Swish}\\
Epochs & 1000–1500\\
Batch size &  $10^4$ \\
Loss function & Negative log-likelihood \\
Learning Rate & \(10^{-4}\)\\
Optimizer & Adam \\
GPU Acceleration & Enabled (if available)\\
Early Stopping & Not applied \\ 
Learning Rate Decay & Not applied \\ 
\hline
\end{tabular}

\end{table}

After training the normalizing flow, the unfolded response is obtained through a Maximum Likelihood Estimation (MLE) step. Here, we minimize the negative log-likelihood (NLL) of the observed data conditioned on the learned model. The MLE process is performed using the same Adam optimizer and can be outlined in Algorithm~\ref{alg:mle_unfolding}.  We did not extensively optimize any of the hyperparameters, but we found that the results were largely insensitive to small changes in the setup.


\begin{algorithm}
\caption{MLE Optimization for Unfolding}
\label{alg:mle_unfolding}
\KwIn{Initial parameters $t_0$, maximum iterations $\text{n\_iter}$, learning rate, tolerance}
\KwOut{Optimized unfolded distribution $t$}
\textbf{Initialize:} $t \gets t_0$\;
\For{$i = 1$ \textbf{to} \text{n\_iter}}{
    Compute $\text{NLL} \gets -\log P(t \mid m; \theta)$ \tcp*{$\theta$ are trained flow parameters}
    Compute gradients $g \gets \nabla_t \text{NLL}$\;
    Update $t \gets \text{AdamUpdate}(t, g, \text{learning\_rate})$\;
    \If{\text{change in NLL} $<$ \text{tolerance}}{
        \textbf{break}\;
    }
}
\Return $t$\;
\end{algorithm}

\section{FBU Implementation}
\label{sec:fbu}

We implement Fully Bayesian Unfolding (FBU) using the Bayesian sampling toolkit PyMC \cite{pymc3}. The setup is the same as for NPU, but we introduce some notation here to connect with the previous literature in the subject.  Given an observed spectrum $\mathbf{d}$ and a response matrix $\mathbf{R}$, the posterior probability of the true spectrum $\mathbf{t}$ follows the probability density
\begin{equation}
\Pr(\mathbf{t}|\mathbf{d},\mathbf{R})
\propto{}
\mathcal{L}(\mathbf{d}|\mathbf{t},\mathbf{R})
\cdot{}
\pi{}(\mathbf{t})
\end{equation}
where $\mathcal{L}(\mathbf{d}|\mathbf{t},\mathbf{R})$ is the likelihood function of $\mathbf{d}$ given $\mathbf{t}$ and $\mathbf{R}$, and $\pi{}$ is the prior probability density for the true spectrum $\mathbf{t}$.
As in Sec~\ref{sec:stats}, assuming a perfect reconstruction efficiency of 1 in all bins,
the simulated detector-level spectrum $\mathbf{m} $ corresponding to a true spectrum $\mathbf{t}$ is given by $m_i(\mathbf{t},\mathbf{R}) = \sum_{j=0}^{N_m}R_{ij}t_j$.
Under the assumption that the data are poissonian counts, the likelihood is then obtained by comparing the observed spectrum
$\mathbf{d}$ with the simulated one:
\begin{equation}
\mathcal{L}(\mathbf{d}|\mathbf{t},\mathbf{R}) =
\prod_{i=1}^{N_m}\text{Poisson}(d_i,m_i(\mathbf{t},\mathbf{R})).
\end{equation}

While the response matrix can be estimated from the simulated sample of signal events, the prior probability density $\pi(\mathbf{t})$ is to be chosen according to what we know about $\mathbf{t}$ before the measurement is performed.
The simplest choice is an uninformative prior that assigns equal probabilities to all $\mathbf{t}$ spectra within a wide range.
In this study, the range $[\mathbf{t}-5\sqrt{\mathbf{t}}, \mathbf{t}+5\sqrt{\mathbf{t}}]$ is considered so that it is sufficiently wide to not introduce biases. A uniform prior is also used for NPU (chosen based on the distribution of parameter points in the training sample) and also is equivalent to a frequentist maximum likelihood estimate.

Having chosen the prior, the posterior probability density $p(\mathbf{t}|\mathbf{d},\mathbf{R})$ is determined by sampling the $N_t$--dimensional parameter space, and evaluating for each point $\mathcal{L}(\mathbf{d}|\mathbf{t},\mathbf{R})$ and $\pi(\mathbf{t})$, thus performing a numerical integration. For a large number of parameters, direct sampling techniques become extremely inefficient; therefore, we use the No-U-Turn Sampler \cite{JMLR:v15:hoffman14a}, an efficient and robust MCMC algorithm.


\section{Numerical Results}
\label{sec:results}

\subsection{2-bin Degenerate Response Example}

\begin{figure}[h!]
\centering
\subfloat[2-bin example]{\includegraphics[width=0.485\textwidth]{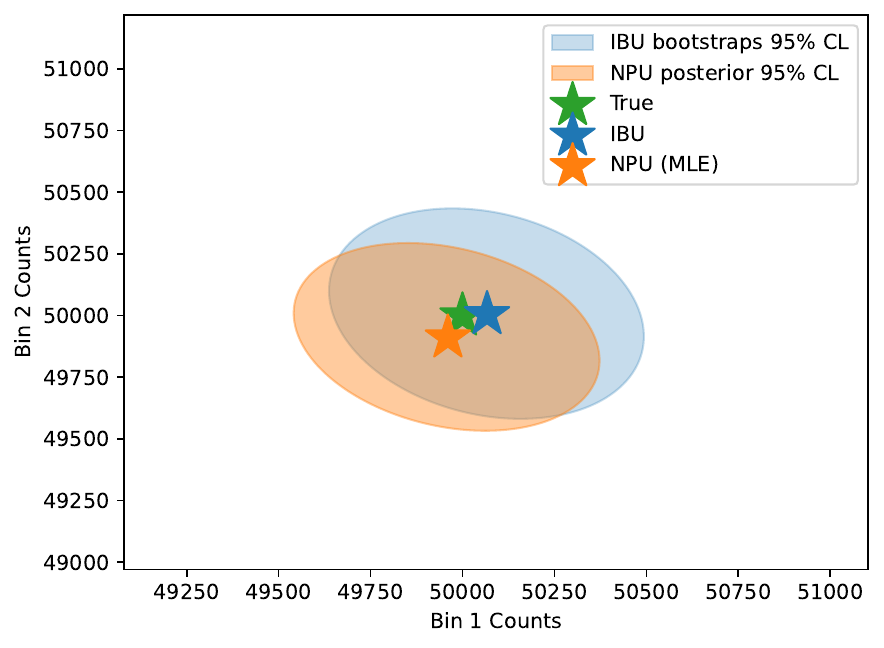}\label{fig:2bin:a}}
\subfloat[2-bin example with degenerate response]{\includegraphics[width=0.49\textwidth]{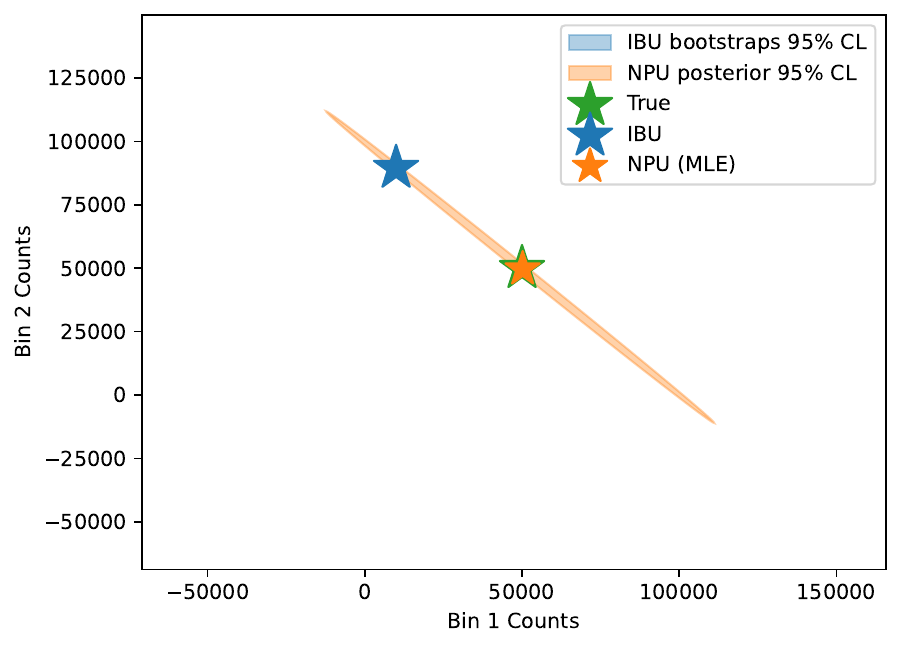}\label{fig:2bin:b}}
\caption{Comparison between NPU and IBU using a two--bin example.
The MLE estimate of NPU is represented with an orange star, the IBU prediction with a blue star, and the true value with a green star. 
The orange and blue shaded regions represent the 95\% credible intervals of NPU and IBU respectively.
(a) A non-degenerate, nearly diagonal response matrix \((\rho = 1)\).
Both IBU and NPU yield unfolded distributions that agree well with the truth.
(b) A degenerate response \((R_{i\alpha} = R_{i\beta})\) leads IBU to rely solely on the prior.
In contrast, NPU produces a full posterior, with credible intervals that properly account for the uncertainty in unconstrained regions.
The  maximum likelihood estimation (MLE) of NPU closely aligns with the truth distributions and the posterior from NPU is compatible with the statistical uncertainty evaluated by IBU using bootstraps.
}
\label{fig:2bin}
\end{figure}

We first illustrate the behavior of NPU using a simple two--bin degenerate example similar to the one introduced earlier.
In this case, $t,m\in\mathbb{R}^2$ and the response matrix $\mathbf{R}\in\mathbb{R}^{2\times 2}$.
We fix $t_0=t_1=5 \times 10^4$ and parameterize $R$ by a correlation coefficient $\rho$ and diagonal elements $\sigma_0=\sigma_1\equiv \sigma$.
In this example, $\sigma=0.8$. 
Fig.~\ref{fig:2bin:a} shows the case where $\rho=0$. 
Since migrations are small, both IBU (via bootstrapping) and NPU yield unfolded confidence regions that are statistically consistent with the truth.
In contrast, as $\rho\to 1$, (see Fig.~\ref{fig:2bin:b}) the detector loses its ability to differentiate between two truth bins.
Here, IBU returns a single value based on the prior, with an (incorrectly) zero uncertainty on the ratio \(\frac{t_\alpha}{t_\alpha+t_\beta}\).
In contrast, NPU naturally returns a broad posterior, with credible intervals that encompass all values consistent with the total counts, thereby appropriately reflecting the degeneracy. 

\subsection{Gaussian Example}

Next, we test the performance of NPU and FBU with a Gaussian example, which also provides insight into the sensitivity of the methods resolution effects.
In this example, each dataset consists of a one--dimensional Gaussian distribution at particle level (with mean \(\mu\) and standard deviation \(\sigma\)) and a corresponding smeared Gaussian at detector level, where the resolution smearing is characterized by a parameter \(\epsilon\).
Two pairs of Gaussian datasets are used.
One pair represents the `simulated' data used to prepare the response matrix and the other pair represents `natural' data with which we check the performance of the unfolding.
Specifically, the first pair of datasets $D_\mathrm{sim}$ is used as the nominal simulation sample, which contains $10^6$ events with $\mu=0$ and $\sigma=1$. 
The second pair of datasets $D_\mathrm{natural}$ is used as the observed data, which contains $10^5$ events with $\mu=0$, $\sigma=1$ and moderately smeared with $\epsilon=0.5$. The structure of the datasets is summarized in Table~\ref{tab:gaussian_datasets}.

\begin{table}[h!]
\centering
\caption{Summary of Gaussian Datasets}
\label{tab:gaussian_datasets}
\begin{tabular}{@{}ccccc@{}}
\toprule
 &  & \(\mu\) & \(\sigma\) & \textbf{N\_events} \\ \midrule
\multirow{2}{*}{Simulation} & Particle-level  & 0       & 1        & \(10^6\) \\ 
                            & Detector-level  & 0       & $\sqrt{1^2+0.5^2}$        & \(10^6\) \\ \midrule
\multirow{2}{*}{Natural}    & Particle-level  & 0       & 1        & \(10^5\) \\
                            & Detector-level  & 0       & $\sqrt{1^2+0.5^2}$ & \(10^5\) \\ \bottomrule
\end{tabular}
\end{table}

\begin{figure}[h!]
\centering
\subfloat[Gaussian setup]{\includegraphics[width=0.435\textwidth]{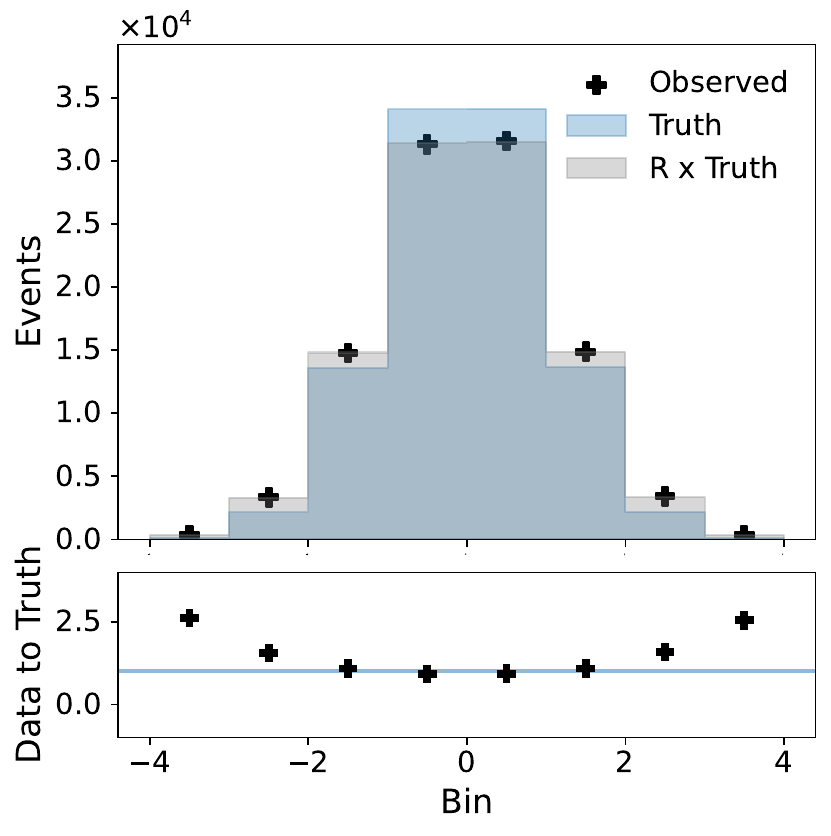}\label{fig:gaus_init:a}} 
\subfloat[Response matrix]{\includegraphics[width=0.53\textwidth]{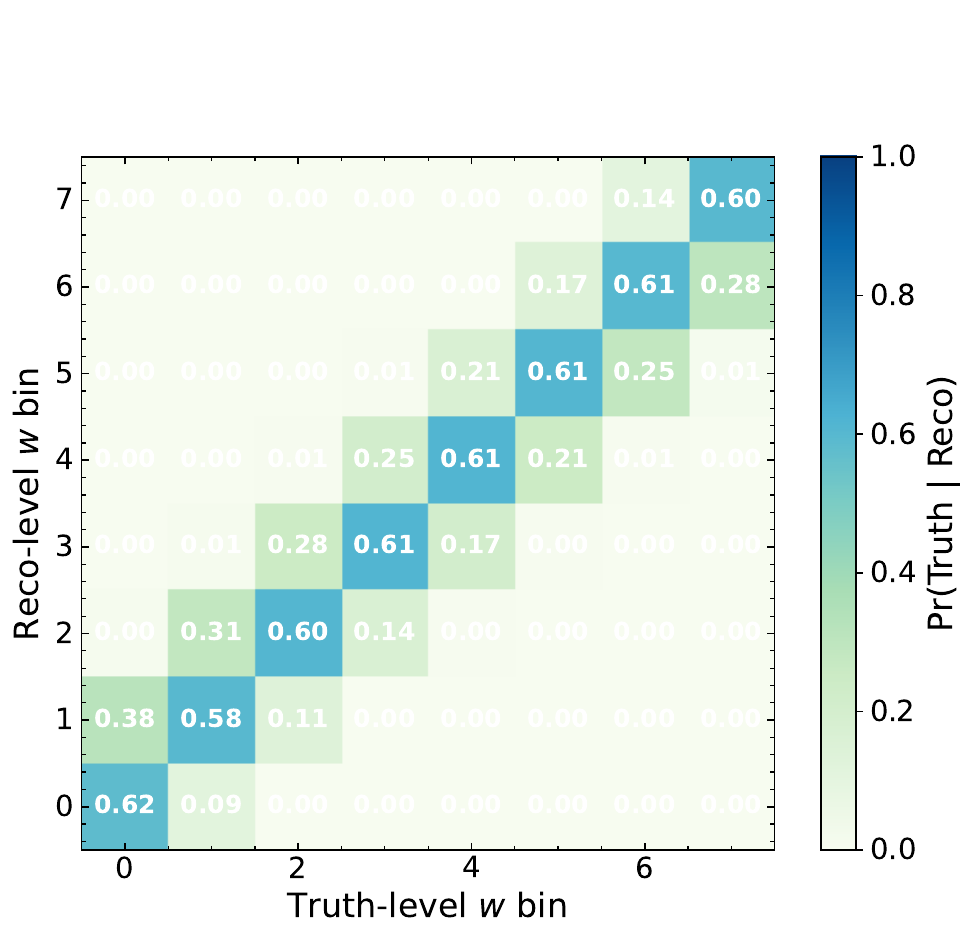}\label{fig:gaus_init:b}} \\
\caption{Gaussian example. (a) the initial setup. The goal of any unfolding method is recover the truth distribution, or the particle-level Gaussian distribution of $D_\mathrm{natural}$ in blue, from the observed (post-detector) distribution of $D_\mathrm{natural}$ here as black crosses. (b) The response matrix ($R$) derived from  $D_\mathrm{sim}$, which captures the smearing and will be used for unfolding.}
\label{fig:gaus_init}
\end{figure}

\begin{figure}[h!]
\centering
\subfloat[Gaussian result]{\includegraphics[width=0.45\textwidth]{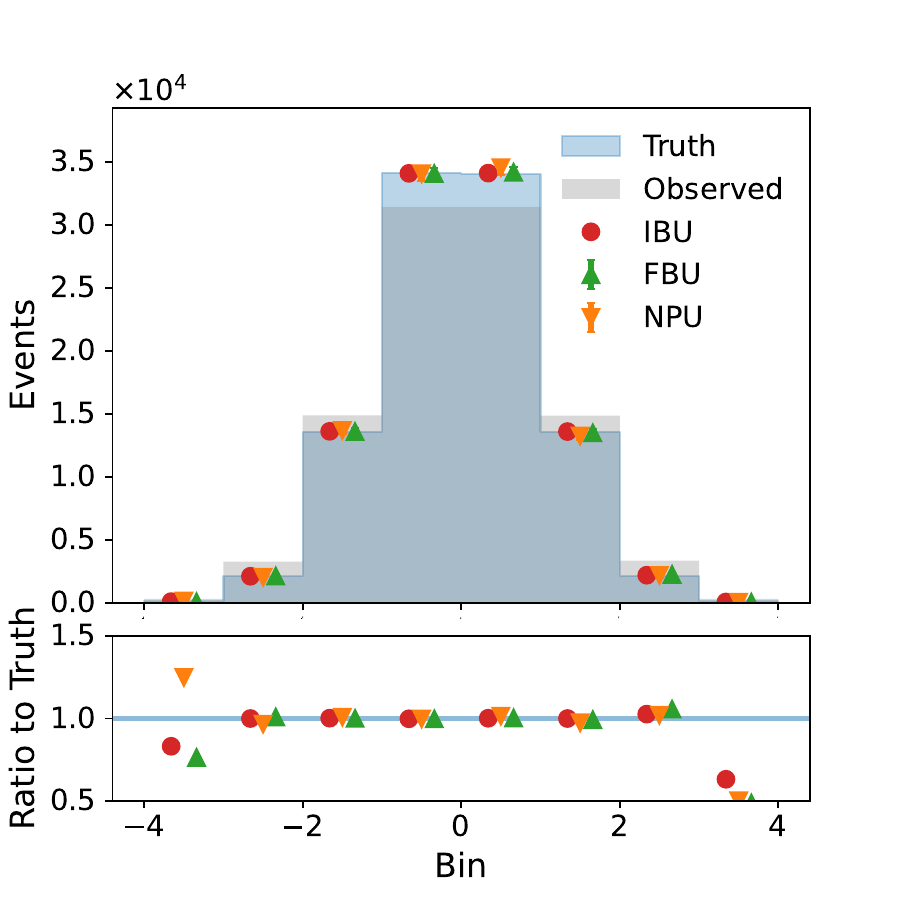}\label{fig:gaus:a}} 
\subfloat[Corner plot for the Gaussian result]{\includegraphics[width=0.44\textwidth]{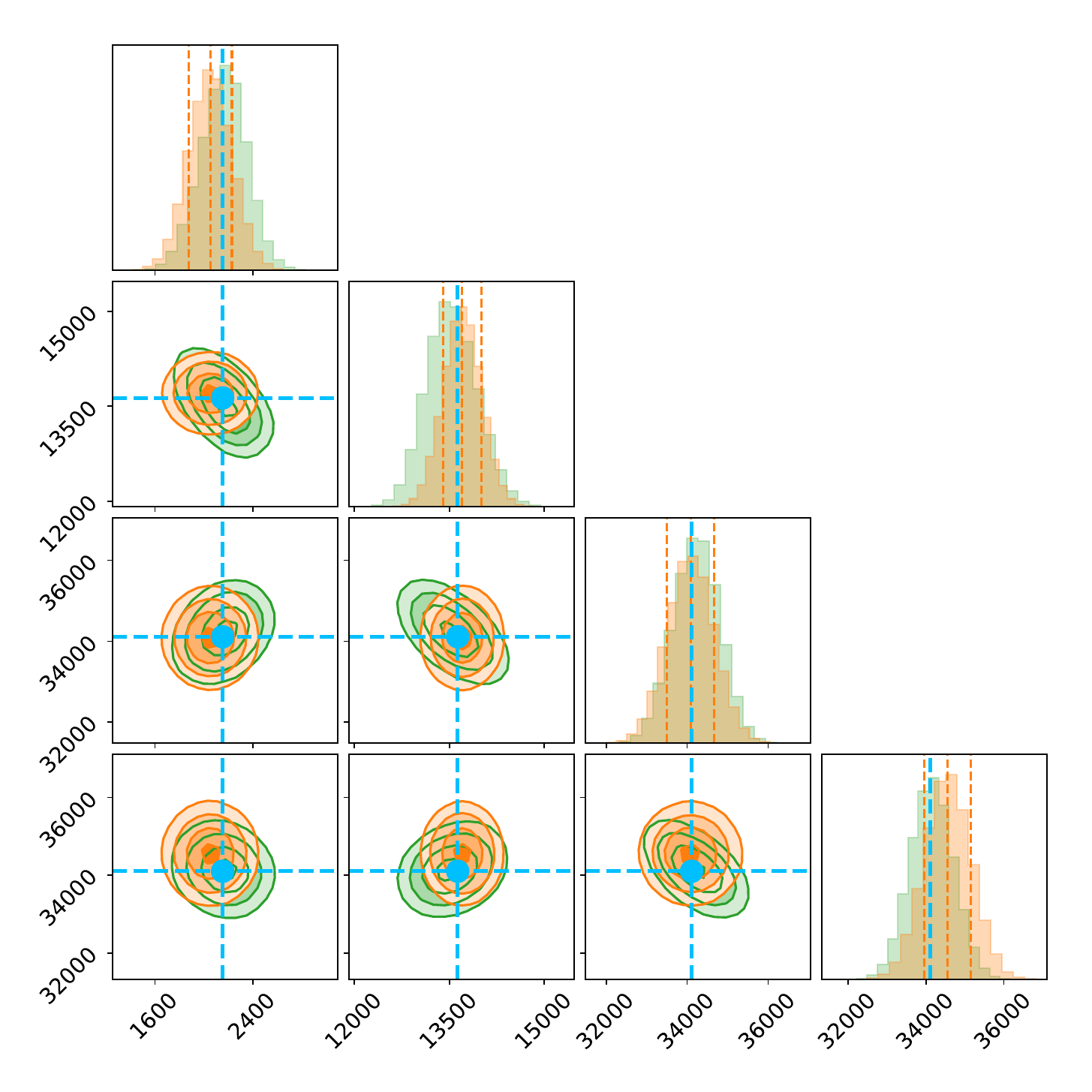}\label{fig:gaus:b}} \\
\caption{(a) The unfolded Gaussian distribution obtained via NPU (MLE as orange down triangles) compared with the truth (blue shaded region), IBU (red circles) and FBU (green up triangles).
The MLE result of NPU shows good agreement with the truth as well the IBU result.
(b) Corner plot showing pairwise correlations between the four central bins in the unfolded posterior from NPU in orange and FBU in green, overlayed with the truth in blue.
%
%
}
\end{figure}

Figure~\ref{fig:gaus_init} illustrates the setup, while Fig.~\ref{fig:gaus:a} shows the marginal unfolded distribution and Fig.~\ref{fig:gaus:b} presents the corresponding corner plot, highlighting the pairwise correlations among the bins.  
Both NPU and FBU accurately and precisely recover the truth, automatically providing the full covariance of the output.  There is some disagreement about the correlations in the posterior between the two methods.  As with the variance itself, mentioned below, this could be a feature of the amortized learning and may be improvable from tuning that step.  

\begin{figure}[h!]
\centering
\subfloat[Pull distribution]{\includegraphics[width=0.445\textwidth]{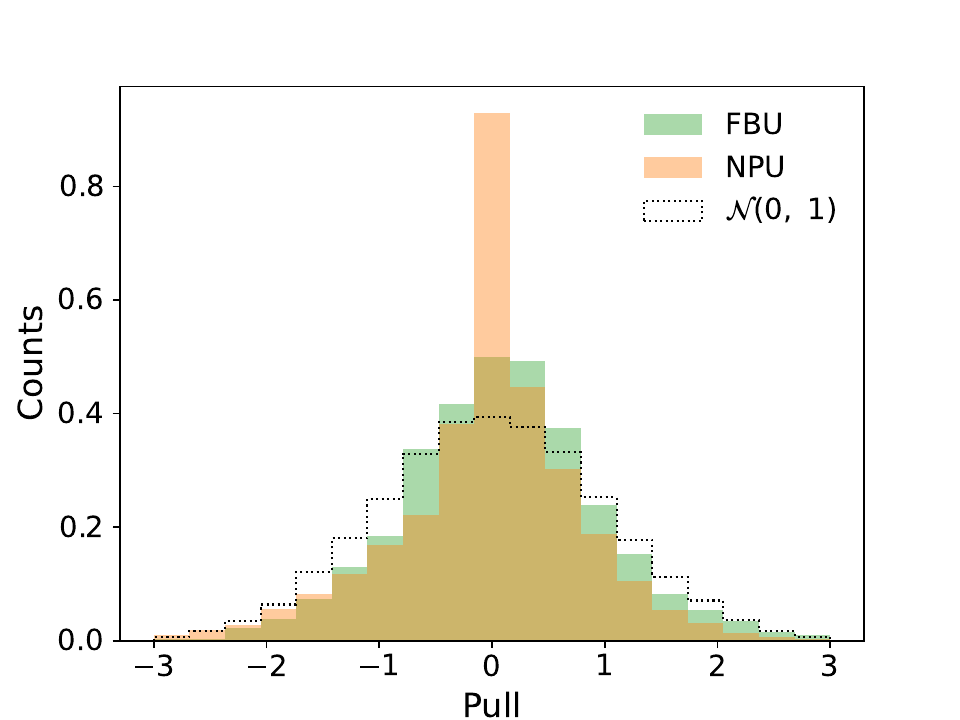}\label{fig:pulls:a}} 
\subfloat[Pulls vs. Smearing Parameter]{\includegraphics[width=0.445\textwidth]{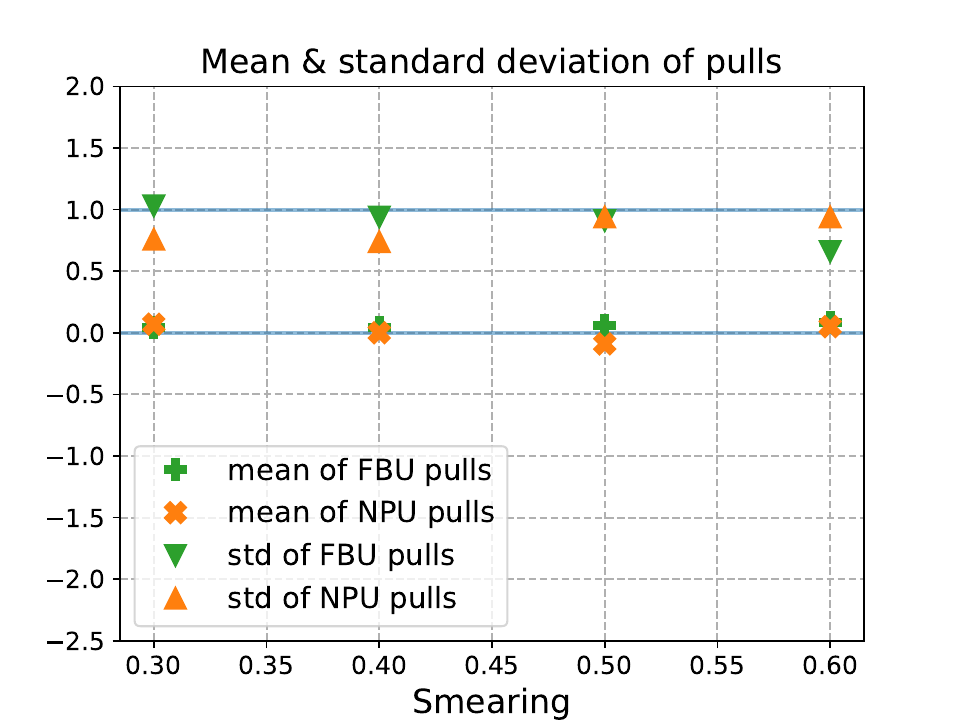}\label{fig:pulls:b}} \\
\caption{Pull distributions from 100 pseudo--experiments for NPU (orange) and FBU (green) in the Gaussian example with \(\epsilon=0.5\), overlaid with a normal distribution \((\mathcal{N}(0, 1))\)  as reference.
 (b) The mean and standard deviation of the pull distributions as functions of the smearing parameter \(\epsilon \in [0.3, 0.6]\). Consistency with \(\mu=0, \sigma=1\) indicates proper uncertainty quantification.
}
\label{fig:pulls}
\end{figure}

We further assess the statistical properties of the results by applying the method to a set of pseudo--experiments via pull distributions, defined as
\begin{equation}
    \operatorname{Pull}_{ij} = \frac{\mu^{\textrm{method}}_{ij} - t_i}{\sigma^{\textrm{method}}_{ij}},
\end{equation}
where method refers to either NPU or FBU, and \(\mu^{\textrm{method}}_{ij}\) and \(\sigma^{\textrm{method}}_{ij}\) refer to the mean and standard deviation of the posterior in bin \(i\) for pseudo--experiment \(j\), respectively. 
Figure~\ref{fig:pulls:a} shows that, for moderate smearing \((\epsilon=0.5),\) the pull distributions for both NPU and FBU are well centered around zero with unit variance, indicating proper calibration. Deviations from \(\mu=0\) and \(\sigma=1\) would suggest model mis-calibration or underestimated uncertainties.
Figure~\ref{fig:pulls:b} displays the dependence of the pull statistics on the smearing parameter \(\epsilon \in [0.3, 0.6]\), confirming that both methods maintain proper coverage across different smearing intensities.
We see that both NPU and FBU return posteriors that continue to meet the expectation of $\mu=0$ and $\sigma=1$ throughout.  The histogram shows that NPU appears to underestimate the variance at low smearing.  As with the posterior correlations described above, this could be improved with additional tweaking of the amortized learning step.

Table~\ref{tab:timing} summarizes the computational cost of performing 100 pseudo--experiments:
FBU with 10k draws and 10k tuning steps takes approximately 40 seconds per pseudo--experiment (totaling 67 minutes), whereas NPU, due to its amortized nature, takes roughly 280 seconds for training and then only a few seconds to process 100 pseudo-experiments.

\vspace{-0.47cm}

\begin{table}[h!]
\centering
\caption{Comparison of computational cost between FBU and NPU. NPU’s amortized inference significantly reduces the runtime for repeated unfolding tasks, despite changes to the data.}
\vspace{3mm}
\label{tab:timing}
\begin{tabular}{|c|c|c|}
\hline
\textbf{Method} & \textbf{Time per Experiment} & \textbf{Total Time (100 Experiments)} \\
\hline
FBU (10k draws) & 40 sec & $\sim$ 67 min (repeated) \\
NPU (training) & 280 sec & $\sim$ 5 min (amortized) \\
\hline
\end{tabular}
\end{table}

\subsection{Particle Physics Example}

We now illustrate the practical application of NPU and FBU through a proof-of-concept study, here focusing on the Large Hadron Collider (LHC).
Specifically, we examine the unfolding performance using two simulated datasets on several widely used jet substructure observables.
Jets are collimated sprays of particles produced by the fragmentation and hadronization of high--energy quarks and gluons.
Ubiquitous and complex, jets serve as an excellent testbed for evaluating unfolding methods, given that detector-induced biases and noise often contribute substantially to experimental systematic uncertainties in jet substructure measurements.
Studies of jet substructure are of interest for diverse LHC programs involving hadronic final states, ranging from stress testing the Standard Model to enhancing search sensitivity for new physics.
We also note that the proposed unfolding methods apply equally well to other collider experiments, including the upcoming Electron Ion Collider (EIC).

Our analysis focuses on jet substructure observables from proton--proton collisions simulated at $\sqrt{s}=14$ TeV, using the same datasets as in Ref.~\cite{Andreassen:2019cjw,andreassen_2019_3548091}.
In our analysis, one simulation (\textsc{Herwig} 7.1.5~\cite{Bellm:2017bvx}) serves as the natural ‘data’ and truth distributions, while another (\textsc{Pythia} 8.243 with Tune 21~\cite{Sjostrand:2014zea, ATL-PHYS-PUB-2014-021}) is used to construct the response matrices. 
To emulate detector effects, we employ the \textsc{Delphes} 3.4.2~\cite{deFavereau:2013fsa} fast simulation of the CMS detector~\cite{CMS:2008xjf}, which incorporates particle flow reconstruction. 
Jets with a radius parameter of $R = 0.4$ are clustered using the anti-$k_T$ algorithm~\cite{Cacciari:2008gp} in FastJet 3.3.2~\cite{Cacciari:2011ma}, applied to either all particle flow objects (detector-level) or stable truth particles excluding neutrinos (particle-level).
To minimize acceptance effects, data is restricted to leading jets in events that contain a $Z$ boson with transverse momentum $p_T^Z > 200$~GeV.
After applying these selection criteria, we retain roughly 1.6 million events from each dataset.

We apply NPU and FBU to several key jet substructure observables~\cite{Larkoski_2020} computed using the datasets as described above, including jet width $\omega$, jet constituent multiplicity $M$, $N$-subjetiness ratio $\tau_{21}=\tau_2^{\beta=1}/\tau_1^{\beta=1}$~\cite{Thaler_2011,Thaler_2012}, and jet momentum fraction $z_g$ after the Soft Drop grooming~\cite{Larkoski_2014, Dasgupta_2013}. 
%
The jet width $\omega$ is the transverse-momentum-weighted first radial moment of the radiation within a jet.  
Gluon initiated jets tend to be wider than quark jets.
Gluon jets also tend to have more constituents than quark jets.

\begin{figure}
\centering
\subfloat[]{\includegraphics[width=0.46\textwidth]{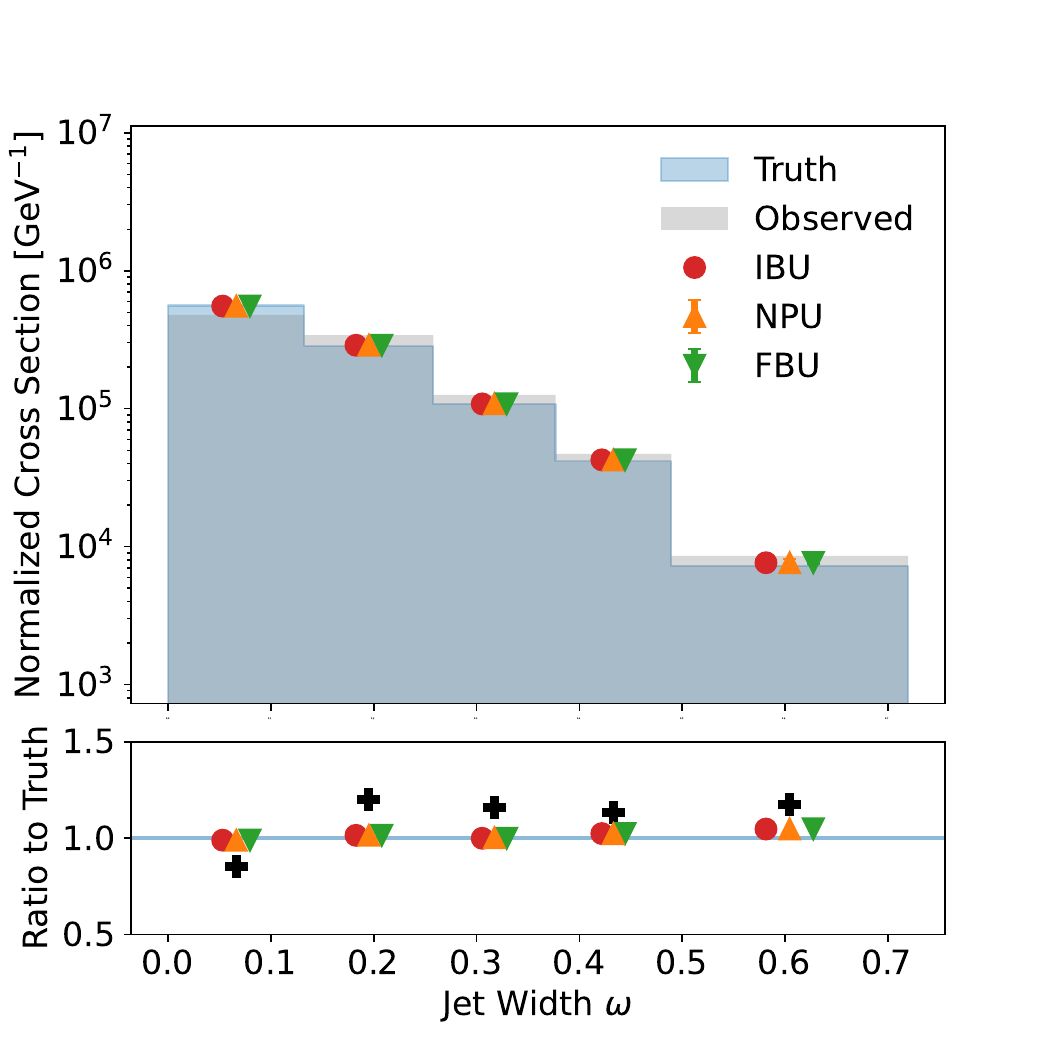}\label{fig:phys_width}} 
\subfloat[]
{\includegraphics[width=0.46\textwidth]{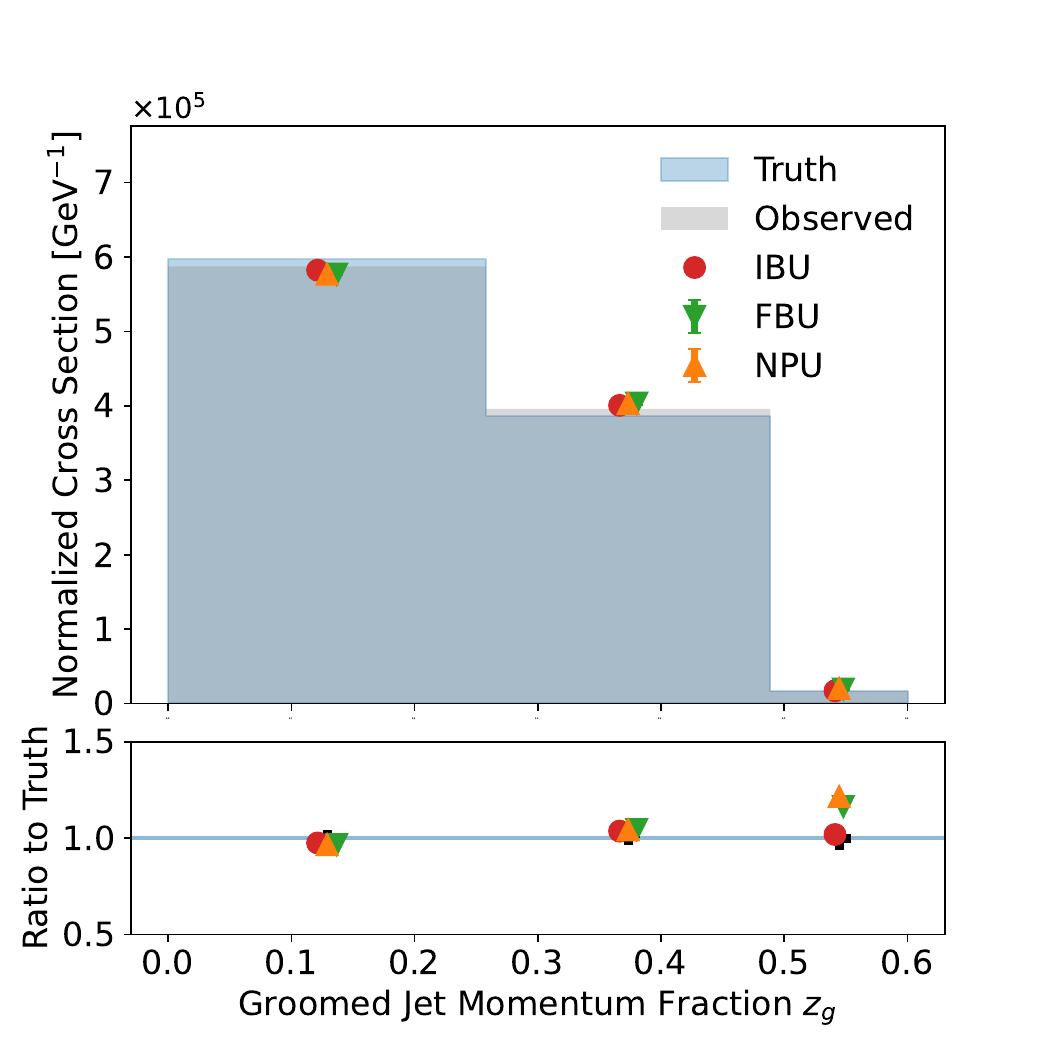}\label{fig:phys_zg}} \\
\subfloat[]
{\includegraphics[width=0.46\textwidth]{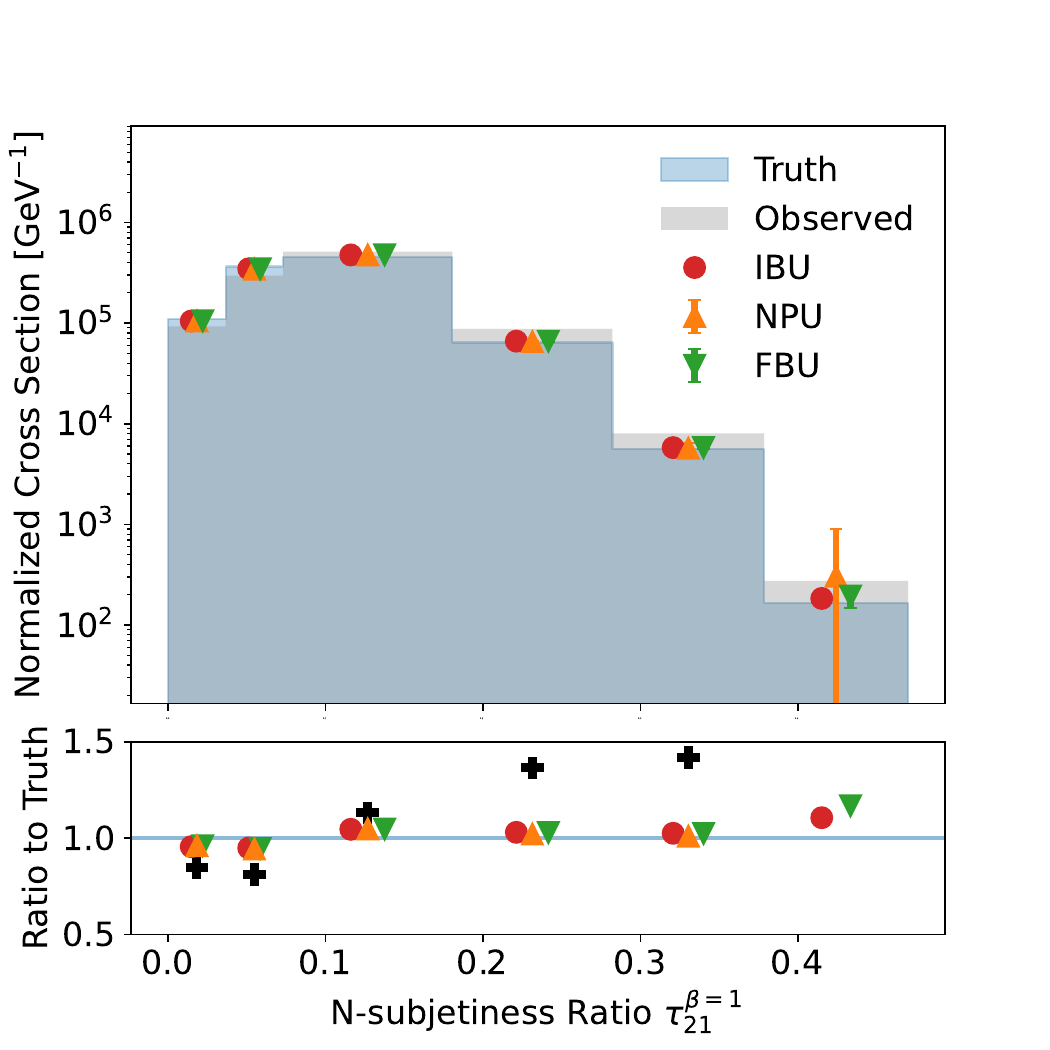}\label{fig:phys_tau}} 
\subfloat[]{\includegraphics[width=0.46\textwidth]{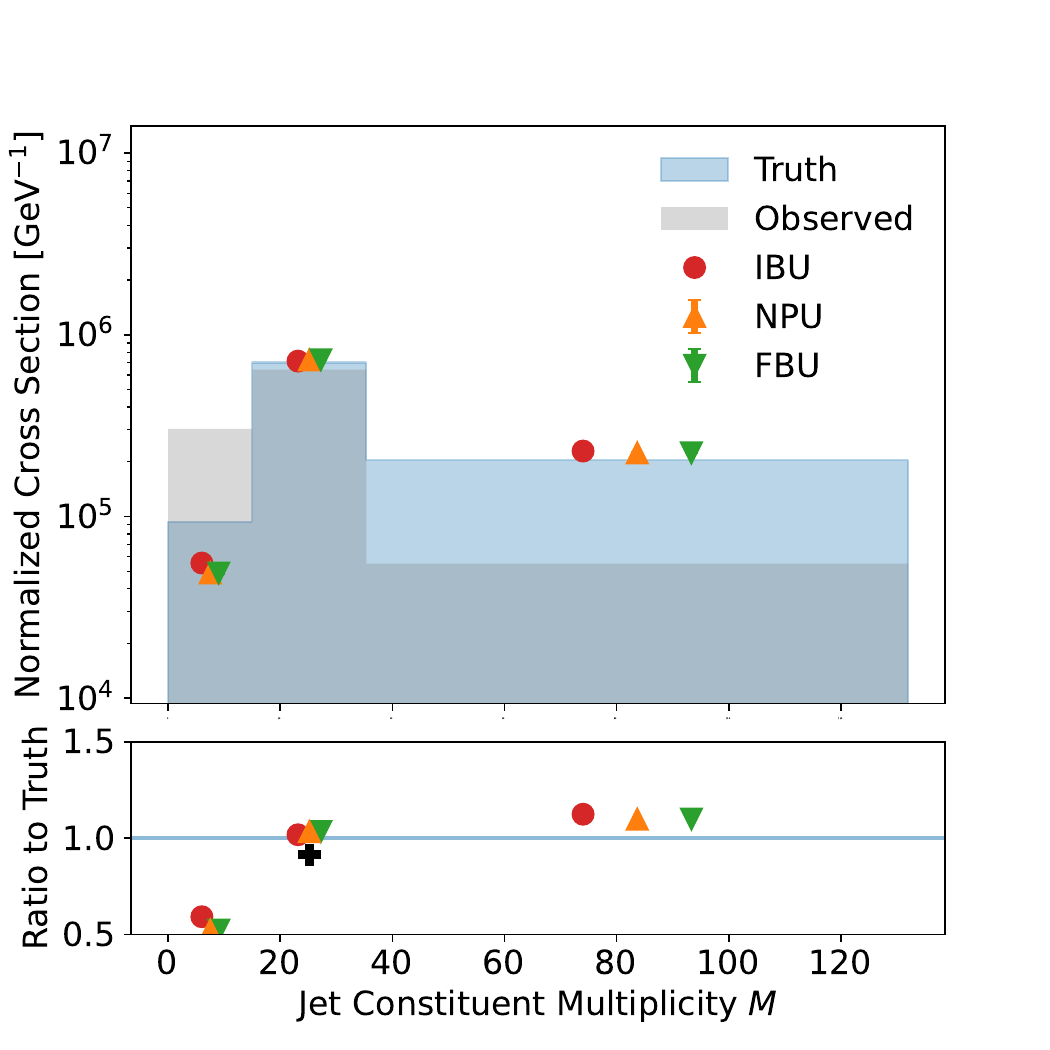}\label{fig:phys_mult}} 

\caption{The unfolding result for (a) jet width $\omega$, (b) jet constituent multiplicity $M$, (c) $N$-subjetiness ratio $\tau_{21}^{\beta=1}$, and (d) groomed jet momentum fraction $z_g$, using \textsc{Herwig} 7.1.5 as data \& truth and \textsc{Pythia} 8.243 Tune 21 for simulations to prepare the response matrix.
The errorbars provide the standard deviation of the posterior from NPU and FBU, in addition to the MLE that continues to succeed in matching the particle level truth distribution (blue shaded).}
\label{fig:substructure}
\end{figure}

Figure~\ref{fig:substructure} demonstrates the unfolded results from NPU and FBU for each of the aforementioned jet substructure observables along with corresponding uncertainty bands, and compares them to the unfolded predictions from IBU with $10 -  15$ iterations.  
%
We note that FBU requires 10x sampling steps to reach convergence in the physics case than for the Gaussian example, namely 100k tuning steps and 500k draws.
The unfolded results mostly agree with each other and the truth, with some disagreements expected from the differences between data and simulation.  In practice, these disagreements would be accounted for as systematic uncertainties.
%
%
In the context of real--world LHC data, where detector effects are intricate and the underlying true distributions may be significantly distorted, the full posterior information provided by NPU is particularly valuable for rigorous uncertainty quantification and subsequent physics interpretation.  In a full analysis, uncertainties in the response matrix itself could also be incorporated, either through repeating the procedure or by including them directly in the likelihood.

\section{Conclusions and Outlook}
\label{sec:conclusions}

In this paper, we proposed Neural Posterior Unfolding (NPU), a generative model-based unfolding method that employs normalizing flows to provide fast access to the full posterior without the need for computationally expensive MCMC sampling through neural posterior estimation\footnote{Unlike cases of computationally expensive forward models, the generative model here is a simple product of Poisson distributions.
However, the computational differences may be non--negligible when many bootstraps are required for classical uncertainty quantification.}.
By directly learning the conditional density $\Pr(t_j \mid m_i)$ from priors processed through the response matrix and obtaining the unfolded result via maximum likelihood estimation, NPU effectively addresses key challenges inherent in classical unfolding methods such as IBU.
In particular, NPU is robust in degenerate cases where traditional methods underestimate uncertainties, as it naturally returns a full posterior that reflects the true level of uncertainty.
See~\cite{Huetsch_2025} for an overview of ML-based unfolding methods.

Our evaluations using a two-bin example verify the coverage of the NPU posterior by comparing it with IBU using bootstraps for a non--degenerate response matrix.
When the response matrix is degenerate, that is to say when the particle level phase space is not uniquely determined by detector level measurements, NPU still yields correct unfolded results, unlike IBU.

We then stress tested the performance of NPU and FBU with a Gaussian example using a range of smearing parameters $\epsilon \in [0.3, 0.6]$ (leading to $50\% - 70\%$ diagonal components in the response matrices). 
This setup involved 8 bins, making it a eight-dimensional unfolding problem for the methods.
Our results confirmed that NPU maintains reasonable performance, returning posteriors that closely aligned with the true distributions through maximum likelihood estimation (MLE).
Additionally, we examined the statistical behavior of NPU and FBU by conducting 100 pseudo-experiments under identical conditions and computing the pull distributions for each bin.
Both methods consistently produced pulls with $\mu=0$ and $\sigma=1$ across the tested smearing parameters $\epsilon \in [0.3,~0.6]$.

In the final part of our study, we employed NPU and FBU to perform cross sections measurements of several jet substructure observables widely used in nuclear and particle physics, including including jet width $\omega$, jet constituent multiplicity $M$, $N$-subjetiness ratio $\tau_{21}$, and jet momentum fraction $z_g$ after the Soft Drop grooming.
We demonstrated that NPU accurately recovered their central values, while also providing access to the full posterior of the unfolded observables.


Unlike classical unfolding methods, the regularization in NPU is implicit.
It would be interesting to explore this difference in more detail in the future in order to understand its benefits and disadvantages.
The NPU framework introduced here may also provide a starting point for full statistical uncertainty quantification in the case of unbinned results.
Lastly, the codes (link below) for NPU and FBU used in this paper are written with modern Python and ML libraries.
Future work will explore the scalability of NPU to higher-dimensional unfolding problems, its sensitivity to prior choices,improving the accuracy of the amortization step and the extension of the framework to unbinned unfolding scenarios. Additionally, integrating alternative generative models could further enhance flexibility and performance.

\section*{Code Availability}

The code for this paper can be found at \url{https://github.com/jp2555/NPU}.

\section*{Acknowledgments}

This work was supported by the U.S. Department of Energy (DOE), Office of Science under contracts DE-AC02-05CH11231 and DE-AC02-76SF00515.

\bibliography{HEPML,other,neurips_2024}
\bibliographystyle{apsrev4-1}

\end{document}